\documentclass[eqsecnum,nofootinbib,floatfix]{revtex4}
\usepackage{amsmath,amssymb,amsfonts,epsfig,graphicx,euscript}%,mathrsfs}

\usepackage{graphics}
\usepackage{float}
\newcommand{\bse}{\begin{subequations}}
\newcommand{\ese}{\end{subequations}}
\newcommand{\be}{\begin{equation}}
\newcommand{\ee}{\end{equation}}
\newcommand{\bea}{\begin{eqnarray}}
\newcommand{\eea}{\end{eqnarray}}
\newcommand{\ba}{\begin{array}}
\newcommand{\ea}{\end{array}}

\begin{document}

\begin{flushright}
\end{flushright}
\begin{flushright}
%{\sf \today}
\end{flushright}
\hfill%
%\vbox{
%    \halign{#\hfil        \cr
%           IPM/P-2010/002\cr
%                     }
%      }
%\vspace{1cm}

\begin{center}

{\LARGE {\sc Laser Interferometer in Presence of Scalar field on Gravitational Wave Background}}

\bigskip
{ Mohammad A. Ganjali\footnote{ganjali@theory.ipm.ac.ir}, Zainab Sedaghatmanesh\footnote{std\_sedaghatmanesh@khu.ac.ir}
} \\
{Department of Physics, Kharazmi University,\\P. O. Box
31979-37551, Tehran, Iran}
\\
\bigskip
\bigskip
\end{center}

\begin{abstract}
Detection of gravitational waves (GW) opened new windows on fundamental physics and it would be natural to search how the role of extra dimensional effects can be traced to gravitational wave physics. In this article, we consider a toy model of five dimensional pure gravity theory compactified on a circle. The resulting four dimensional theory is a scalar-Maxwell theory which is minimally coupled with gravity. By finding the equations of motion for scalar, electric and magnetic fields, we would be able to find exact wave solutions of coupled equations which are zero mode solutions. We also perform perturbation in order to consider non-zero modes of electromagnetic fields.\\
Having these solutions at hand, we study the recombination of scalar-affected electromagnetic waves (EWs) in a typical Michelson interferometer. In particular, we obtain, up to first order, the change of amplitude of electromagnetic power due to presence of this scalar field which may reveal some signals of extra dimension.
\end{abstract}

\maketitle

\section{Introduction}
\label{intro}
A century after the presentation of General Relativity (GR) by Einstein, one of the most important predictions of this theory was confirmed through direct detection of Gravitational Waves (GWs) by Laser Interferometer GW Observatory (LIGO) experiment \cite{Abbott:2016blz}. The observed signal in this observatory shows that the GW comes from the quasi-circular inspiral, merger and ringdown of binary black holes. The signal-to-noise ratio was 24 for this GW150914 with statistical $\sigma>5$. In fact LIGO with strain sensitivity of order $10^{-21}$ showed the proximity of the source to Earth $(420^{+150}_{-180}Mpc)$ and the mass of the binary $(m_1,m_2)=(36^{+5}_{-4}, 29^{+4}_{-4})M_{\odot}$ \cite{TheLIGOScientific:2014jea}-\cite{TheLIGOScientific:2016pea}. Soon after the first LIGO test, it's results were confirmed by the detection of GW151226 event\cite{Abbott:2016nmj}.\\
Focusing on part of the sky from where the LIGO received GWs, another fascinating observation was made by the Fermi Gamma-ray Space Telescope in which the existence of a weak gamma-ray burst above $50 KeV$ just $0.4s$ after GW150914 was reported and its position overlapped with that of LIGO's observation \cite{Connaughton:2016umz}.

After that, LIGO/VIRGO consortium announced the merger of two neutron stars, GW170817, in Galaxy NNGC4993 about $40Mpc$ from the Milky Way \cite{TheLIGOScientific:2017qsa}\footnote{ After the events GW150914 \cite{Abbott:2016blz} and GW151226 \cite{Abbott:2016nmj}, the LIGO/VIRGO announced other GW detections like GW170104 \cite{Abbott:2017vtc}, GW170814 \cite{Abbott:2017oio}, GW170817 \cite{TheLIGOScientific:2017qsa} , GW170608 \cite{Abbott:2017gyy} ...GW190521\cite{GW190521}}. Again an electromagnetic counterpart (i.e., the gamma-ray burst GRB170817A) was detected by Fermi gamma-ray telescope \cite{Monitor:2017mdv} and other telescopes \cite{GBM:2017lvd} with the observation that the speed of two waves (i.e., gravitational and electromagnetic waves (EW)) can differ $\frac{c_g^2-c^2}{c^2}\leq6\times 10^{-15}$ where $c_g$ is the speed of GW and $c$ is the speed of light.

The detection of GWs opens new windows on fundamental physics. It is believed that radiated waves from various sources give us rich information about astrophysical objects, early universe, cosmic microwave background, and the like and specifically quantum gravity (\cite{Konoplya:2016pmh} and references therein). See also \cite{Aldrovandi:2008ci} for a review. Several studies with different approaches have been done in order to investigate the presence of electromagnetic radiation associated with GW, see for example \cite{Li:2016zjc}-\cite{Andriot:2017oaz}. In this respect, some efforts have been done to investigate the physical effects of gravity-electromagnetic interactions. In fact, Maxwell equations in the presence of non-zero off-diagonal components of GW metric may cause some new interesting electromagnetic phenomena. For example, gravito-magnetic effect on gyroscopes due to Earth's rotation has been measured\cite{Everitt:2011hp}. For more studies in gravito-magnetic theories see for example \cite{Tajmar:2004ww}.

The question that naturally arises is whether higher dimensional effects can be seen on GWs? Many attempts were made to answer this question by taking into account various theories in higher dimensions \cite{Hogan:2000is}-\cite{Khalil:2018aaj}. In most of these theories, in reduction to four dimensions, even if we start with a pure geometric theory, we will have a gauge field in addition to the four dimensional metric field, and this is  good news because, the higher dimension can be considered as the source of the electromagnetic counterpart of GWs. We may also have additional massless or massive scalar, vector or tensor fields in this theory. The resultant theory would be an effective interacting theory between gravity, electromagnetism, and some additional fields available due to the dimensional reduction.
Therefore, studying theories about the interaction between gravity, U(1) gauge field and some other fields due to higher dimensions may give us signatures of higher dimensions.

Given the above facts and noting that the LIGO experiment works based on the interference of two light waves passing two different optical paths due to the gravitational effect (gravitational path-difference), we would like to study a four dimensional minimal interacting scalar-Maxwell theory on the background of a GW in order to explicitly demonstrate the effects of these scalar-gauge and gravity-gauge interactions on the phase differences of two light waves in a LIGO-like experiment\footnote{See \cite{Blaut:2015qaa},\cite{Cabral:2016klm} for works with similar ideas.}.
We will obtain corrections to the phase of light waves in terms of the scalar field and GW parameters and discuss their possible effects on the light interferences.

At the end of this section, it is noteworthy that our starting point is five dimensional pure gravity. Further, compactifying one extra dimension would provide a four dimensional scalar-Maxwell theory which is minimally coupled to gravity. However, it would also be interesting to consider the generation of GW in the five-dimensional world from the merger of two neutron stars or black holes and the like and to study its effect on the interference of two EWs by using a laser interferometer.

This paper is organized as follows: After a quick review of the compactification of a pure gravity theory from five dimensions to four dimensions in section II,  we will derive the equations of motion in section IIA and try to solve them to find some simple solutions and also  wave solutions in sections IIB and IIC respectively. In section III, we will apply the solutions for a LIGO-like experiment to study the direct effects of scalar fields and GWs on EWs and discuss possible experimental signals of higher dimensions. In section IV, we will end up the paper with concluding remarks.
\section{From Five to Four}
\label{sec:1}
In this section, first, we quickly review the reduction of a five-dimensional pure gravity theory to four dimensions. Such a reduction can be done by compactifying one extra dimension. Due to this compactification one gets massless tensor, vector and scalar modes and the corresponding Lagrangian would be an interacting Lagrangian between these modes. Varying the action with respect to various fields would provide corresponding equations of motion.

Eventually, we will solve the equations of motion for scalar and gauge fields on the background of a GW at the linear level. We will perform this procedure without considering the back-reaction of the presence of scalar and gauge fields on GW dynamics.
\subsection{Scalar-Maxwell-Gravity Interaction from Extra Dimension}
\label{sec:2}
Here, we will consider the simplest version of a five-dimensional pure gravity theory which was originally proposed by Kaluza(1921) \cite{Kaluza:1984ws}. Compactifying one extra dimension on a circle would give us an effective theory in four dimensions which includes scalar, vector, and tensor fields.
In particular, the story begins by writing the five-dimensional metric as \cite{Overduin:1998pn}
\bea\label{metric5}
g_{\hat{\mu}\hat{\nu}}=
\begin{pmatrix}
g_{\mu\nu}-\varphi A_{\mu}A_{\nu}& -\varphi A_{\mu} \\
-\varphi A_{\nu}& -\varphi
\end{pmatrix}
\eea
where hat indicates the five-dimensional variable, $x^{\hat{\mu}}=(x^{\mu},y),\;\mu=0,1,2,3$. Then, considering independence of fields from the fifth coordinate $y$, by an infinitesimal coordinate transformation as $x^{\mu}\rightarrow x^{\mu}+\epsilon\xi^{\hat{\mu}}(x^{\mu})$ the above fields would change as
\bea
\delta\varphi&=&\xi^{\rho}\partial_{\rho}\varphi,\cr
\delta A_{\mu}&=&A_{\rho}(\partial_{\mu}\xi^{\rho})+\xi^{\rho}(\partial_{\rho}A_{\mu})+\partial_{\mu}\xi^5,\cr
\delta g_{\mu\nu}&=&g_{\mu\rho}(\partial_{\nu}\xi^{\rho})+g_{\rho\nu}(\partial_{\mu}\xi^{\rho})
+\xi^{\rho}(\partial_{\rho}g_{\mu\nu}).
\eea
These transformations are the same transformation rules that one expects for scalar, vector, and symmetric rank-2 tensor fields in four dimensions. Next, writing the five-dimensional action as
$$S^{(5)}=-\int{d^5x\sqrt{-\hat{g}}\hat{R}}$$
and compactifying the fifth dimension $y$ on a circle with radius $R_0$ in which $y\sim y+2\pi R_0$ and redefining the scalar field $\varphi$ and field strength $F_{\mu\nu}=\partial_{\mu}A_{\nu}-\partial_{\nu}A_{\mu}$ as $\varphi'=\sqrt{3}\varphi,\;\; F^2=\varphi^{-1}F'^2$,
the following action is obtained in a familiar form as
\bea\label{action}
S=\int{d^4x\sqrt{-g}\left\{-R+\frac{1}{2}\partial_{\mu}\varphi\partial^{\mu}\varphi-
\frac{1}{4}e^{-\sqrt{3}\varphi}F_{\mu\nu}F^{\mu\nu}\right\}},
\eea
where the prime was omitted in the final expression of $S$ and $g=det(g_{\mu\nu})$.

It is notable that the effect of the fifth dimension is encoded on the dynamics of the scalar field $\varphi$. In particular, we know that the radius of the fifth dimension $R_0$ is related to $\varphi$ as
\bea\label{R}
R_0\propto (-\varphi)^{\frac{1}{2}}
\eea
Our aim in this paper is to investigate this theory by deriving the equations of motion of various fields and seek to solve them on the background of a GW and then demonstrate the effects of scalar and metric fields on the phase difference of two traveling lights in a LIGO-like experiment. We will do our analysis at the linear level of field perturbations.
\subsection{Equations of Motion}
\label{sec:3}
The equations of motion for dynamical fields $\varphi, A_{\mu}$ and $g_{\mu\nu}$ derived from the action (\ref{action}) read as
\bea
\label{einstein}
&R_{\mu\nu}-\frac{1}{2}g_{\mu\nu}R-\partial_{\mu}\varphi\partial_\nu\varphi+
e^{-\sqrt{3}\varphi}g^{\alpha\beta}F_{\mu\alpha}F_{\nu\beta} \nonumber\\
&+g_{\mu\nu}\left[\frac{1}{2}\partial_{\alpha} \varphi \partial^{\alpha}\varphi-\frac{1}{4}e^{-\sqrt{3}\varphi}F_{\alpha\beta}F^{\alpha\beta}\right]=0,\\
\label{maxwell}
&\frac{1}{\sqrt{-g}}\partial_{\nu}(\sqrt{-g}e^{-\sqrt{3}\varphi}F^{\mu\nu})=0 ,\\
\label{motion,fi}
&\frac{1}{\sqrt{-g}}\partial_\mu (\sqrt{-g}\partial^\mu \varphi) -\frac{\sqrt{3}}{4}e^{-\sqrt{3}\varphi}F_{\mu\nu}F^{\mu\nu}=0,
\eea
%Now the problem is to solve these equations for our three variables $A_\mu,\varphi$ and $g_{\mu\nu}$.
Our aim is to study the above generalized scalar-Maxwell equations on a GW background.
\subsubsection{Einstein equation}
As we mentioned, we want to find the wave solutions of the first equation (\ref{einstein}) which is just the Einstein equation with sources. As usual, one inserts flat space metric perturbations $g_{\mu\nu}=\eta_{\mu\nu}+h_{\mu\nu}$ such that $\vert h_{\mu\nu}\vert\ll1$ in (\ref{einstein}) and obtains linearised equation for the metric field in the presence of source fields $\varphi$ and $F_{\mu\nu}$. This procedure should consistently be accompanied by solving (\ref{maxwell}) and (\ref{motion,fi}). Using the standard gauge fixing prescription in four dimensions one can see that the linearized version of (\ref{einstein}) reads as
\bea\label{leinstein}
\square h_{\mu\nu}=-2T_{\mu\nu}
%-\partial_{\mu}\varphi\partial_\nu\varphi&+&
%e^{-\sqrt{3}\varphi}(\eta^{\alpha\beta}+h^{\alpha\beta})F_{\mu\alpha}F_{\nu\beta}
%+\frac{1}{2}\left(\eta_{\mu\nu}\eta^{\alpha\beta}+\eta_{\mu\nu}h^{\alpha\beta}
%+h_{\mu\nu}\eta^{\alpha\beta}\right)\partial_{\alpha} \varphi \partial_{\beta}\varphi\nonumber\\
%&-&\frac{1}{4}e^{-\sqrt{3}\varphi}\left(\eta_{\mu\nu}\eta^{\alpha\gamma}\eta^{\beta\lambda}
%+\eta_{\mu\nu}\eta^{\alpha\gamma}h^{\beta\lambda}
%+\eta_{\mu\nu}h^{\alpha\gamma}\eta^{\beta\lambda}
%+h_{\mu\nu}\eta^{\alpha\gamma}\eta^{\beta\lambda}\right)
%F_{\alpha\beta}F_{\gamma\lambda}=0.
\eea
where $\square$ is the flat space Da'lamberian operator and $T_{\mu\nu}$ is the energy-momentum tensor evaluated up to the linear order of $h_{\mu\nu}$.
Moreover, we would like to have a plane wave solution of (\ref{leinstein}) propagating in the $z$ direction. In particular, we want to consider GW of the form $g_{\mu\nu}=g_{\mu\nu}(z-ct)$. Due to this assumption, i.e. the dependence of the metric on $z-ct$, and using the standard gauge fixing prescription in four dimensions, one can extract dynamical propagating modes which are $h_{xx}=-h_{yy}$ and $h_{xy}$ and show that the metric would take the following form
\bea
\label{metric}
ds^2&=&c^2dt^2-dz^2-\left(1- f_{+}\right)dx^2-\left(1+ f_{+}\right)dy^2
+2f_{\times}dxdy ,\nonumber
\eea
Thus, it remains to examine that the metric (\ref{metric}) is a solution of (\ref{leinstein}). Since $h_{xx}=-h_{yy}$ and $h_{xy}$ are functions of $z-ct$ thus, $\square h_{ij}=0,\;\;{i,j}={x,y}$. However, this is incomplete because of the presence of perturbative source terms in (\ref{leinstein}). Consequently, the presence of source terms would add perturbative terms to the plane wave metric (\ref{metric}). Here, we stop analysing (\ref{leinstein}) and will return to it when finding the explicit solutions of full equations of motion (\ref{einstein}-\ref{motion,fi}).
\subsubsection{Maxwell equation}
The second equation together with the Bianchi identity for $F_{\mu\nu}$
\bea\label{bianchi}
\partial_{[\rho}F_{\mu\nu]}=0,
\eea
 are actually the generalized Maxwell equations in four-dimensional curved space-time which is now accompanied by $\varphi$. It would be interesting to note that the first equation (\ref{maxwell}) can be rewritten as
\begin{align}
\label{maxwelll}
\frac{1}{\sqrt{-g}}\partial_{\nu}(\sqrt{-g}F^{\mu\nu})=J^{\mu},
\end{align}
where
\bea\label{current}
J^{\mu}=\sqrt{3}\partial_{\nu}\varphi F^{\mu\nu},
\eea
is the induced current due to the presence of the interaction between the scalar field $\varphi$ and the electromagnetic field. Therefore, dynamical scalar field would imply induced charge and current density in this theory.

Introducing the following definitions
\begin{align}
F_{0k}\equiv\dfrac{E_k}{c},\;\;\;\;\;\;
F_{jk}\equiv -\epsilon_{ijk}B^i,
\end{align}
then, we can rewrite the equations (\ref{maxwell}) for $\mu=0$ and $\mu=i$ and get the generalized Gauss and Maxwell-Ampere laws as
\begin{align}
\label{gauss}
&\partial_k  E_j (g^{0k}g^{j0}-g^{jk}g^{00})-\partial_\mu B^k cg^{m\mu}g^{n0}\varepsilon_{kmn}+E_j \gamma^j \nonumber\\
& -B^k c\varepsilon_{kmn}\sigma^{mn0}-\sqrt{3}\big[(g^{0\mu}g^{0j}-g^{j\mu}g^{00})\dfrac{E_j}{c}\nonumber\\
 &-g^{m\mu}g^{n0}\varepsilon_{kmn}B^k\big]\partial_\mu \varphi=0 ,
\end{align}
\begin{align}
\label{ampere}
&\dfrac{1}{c}\partial_\mu E_j(g^{0\mu}g^{ji}-g^{j\mu}g^{0i})-\varepsilon_{kmn}\partial_\mu B^kg^{m\mu}g^{ni}+E_j\xi^{ij}\nonumber\\
&-B^k \varepsilon_{kmn}\sigma^{mni}-\sqrt{3}\big[(g^{0\mu}g^{ji}-g^{j\mu}g^{0i})\dfrac{E_j}{c}\nonumber\\
&-g^{m\mu}g^{ni}\varepsilon_{kmn}B^k\big]\partial_\mu\varphi =0 ,
\end{align}
where $\gamma^j, \sigma^{mn\beta}, \xi^{ji}$  are defined as follows \cite{Cabral:2016klm}
\bea
\gamma^j&\equiv &\big{[} \partial_k(g^{0k}g^{j0}-g^{jk}g^{00}) \nonumber \\
&+&\frac{1}{\sqrt{-g}}\partial_k(\sqrt{-g})(g^{0k}g^{j0}-g^{jk}g^{00})\big{]}
\eea
\bea
\xi^{ij} &\equiv &\frac{1}{c}\big{[} \partial_{\mu}(g^{0\mu}g^{ji}-g^{j\mu}g^{0i}) \nonumber \\
&+&\frac{1}{\sqrt{-g}}\partial_{\mu}(\sqrt{-g})(g^{0\mu}g^{ji}-g^{j\mu}g^{0i})\big{]}
\eea
\bea
\sigma^{mn\beta}\equiv \big{[} \partial_{\mu}(g^{m\mu}g^{n\beta})
+\frac{1}{\sqrt{-g}}\partial_{\mu}(\sqrt{-g})(g^{m\mu}g^{n\beta})\big{]}\nonumber\\
\eea
We also have Faraday and magnetic Gauss laws as
\begin{align}\label{faraday}
\partial_x B^x+\partial_y B^y+\partial_z B^z=0,\nonumber\\
\partial_tB^x +\partial_y E_z-\partial_z E_y=0,\nonumber\\
\partial_t B^y+\partial_z E_x-\partial_x E_z=0,\nonumber\\
\partial_t B^z+\partial_x E_y-\partial_y E_x=0.
\end{align}
One can see that the presence of the off-diagonal components of the metric in Ampere and Gauss laws would cause new effects on electromagnetic phenomena.

Indeed, by a glimpse on (\ref{maxwell}) or noting that factor $(g^{0\mu}g^{0j}-g^{j\mu}g^{00})$ in the second line of (\ref{gauss}) is zero for $\mu=0$ then, Gauss and Ampere equations, (\ref{gauss}) and (\ref{ampere}), can be rewritten as
\begin{align}
\label{gausss}
\partial_k  \hat{E}_j (g^{0k}g^{j0}-g^{jk}g^{00})&-\partial_\mu \hat{B}^k cg^{m\mu}g^{n0}\varepsilon_{kmn}
\\
&+\hat{E}_j \gamma^j
-\hat{B}^k c\varepsilon_{kmn}\sigma^{mn0}=0 ,\nonumber
\end{align}
\begin{align}
\label{amperee}
\dfrac{1}{c}\partial_\mu \hat{E}_j(g^{0\mu}g^{ji}-g^{j\mu}g^{0i})&-\varepsilon_{kmn}\partial_\mu \hat{B}^kg^{m\mu}g^{ni}\\
&+\hat{E}_j\xi^{ij}
-\hat{B}^k \varepsilon_{kmn}\sigma^{mni}=0 ,\nonumber
\end{align}
where in the presence of the field $ \varphi $ we have defined 
\begin{align}
\hat{E}_j\equiv e^{-\sqrt{3}\varphi}E_j,\;\;\;\;\;
\hat{B}^{j}\equiv e^{-\sqrt{3}\varphi}B^j
\end{align}
%In order to solve the above equations (\ref{maxwell})-(\ref{einstein}), the typical method is to consider these coupled equations together and try to solve them simoltaneously. But, as it may result in numerical solutions we prefer to assume that
Considering $f_{\times}=0$ for simplicity, and applying metric (\ref{metric}) to equations (\ref{gauss} and \ref{ampere}),
%and taking in to account $\varphi=\varphi(t,z)$ then,
we get\footnote{Hereafter, up to section III, we will set c=1}
\bea\label{amp-gauss}
&&g^{xx}\partial_x \hat{E}_x +g^{yy}\partial_y \hat{E}_y-\partial_z \hat{E}_z-\hat{E}_z\gamma^z =0,\nonumber\\
&&g^{xx}\partial_t \hat{E}_x+\xi^{xx}\hat{E}_x%\nonumber\\
+g^{xx}g^{yy}\partial_y \hat{B}^z+g^{xx}\partial_z \hat{B}^y-\hat{B}^y\sigma^{zxx}=0,\nonumber\\
&&g^{yy}\partial_t \hat{E}_y +\xi^{yy}\hat{E}_y%\nonumber\\
-g^{yy}g^{xx}\partial_x \hat{B}^z-g^{yy}\partial_z \hat{B}^x
+\hat{B}^x\sigma^{zyy}=0,\nonumber\\
&&g^{zz}\partial_t \hat{E}_z +\xi^{zz}\hat{E}_z-g^{xx}\partial_x \hat{B}^y+g^{yy}\partial_y \hat{B}^x=0,
\eea
Moreover, following \cite{Cabral:2016klm}, given that the metric does not depend on $x$ and $y$ coordinates , equations (\ref{gauss}) and (\ref{ampere}) in the presence of the external source field $j^{\mu}=(\rho, \vec{J_0})$, can be rewritten in the familiar form as
\begin{align}
\partial_k \widetilde{E}^{k}&=\dfrac{\varrho}{\varepsilon_0}\\
\epsilon_{ijk}\partial_{j}\overline{B}^{iijjk}&=\mu_{0}(j_0^i+j^i_D).
\end{align}
where we have defined
\begin{align}
\widetilde{E}^j&\equiv -g^{jj}g^{00}\sqrt{-g}e^{-\sqrt{3}\varphi}E_j\\
\overline{B}^{iijjk}&\equiv g^{ii}g^{jj}\sqrt{-g}e^{-\sqrt{3}\varphi}B^k
\end{align}
and
\begin{align}
\varrho &\equiv \sqrt{-g}\rho\\
j^i&\equiv \sqrt{-g}J_0^i\\
j^i_D&\equiv -\varepsilon_0 \partial_t (\sqrt{-g}g^{00}g^{ii} e^{-\sqrt{3}\varphi} E_i)
\end{align}
In all the above expressions no summation convention is assumed for index $i$. In the above relation, $j^i_D$ denotes the usual "Displacement Current". Note that despite the usual flat space electromagnetism, the displacement current is non-zero for the constant electric field on the time dependent background $g_{\mu\nu}$.
\subsubsection{Scalar equation}
As $\varphi$ represents the fifth dimension and due to symmetry considerations, we may naturally expect that it does not depend on the spatial components of the space-time but if we take a gravitational wave propagating in the z direction, the rotational symmetry of space-time is broken in the z direction and consequently, we take $\varphi$ as a function of $(t,z)$. Thus, the equation of motion for the scalar field reads as
\bea
\label{fi,second}
\partial^2_t \varphi -\partial^2_z \varphi &+&\dfrac{\partial_t(\sqrt{-g})}{\sqrt{-g}}(\partial_t\varphi +\partial_z\varphi)\\
&-&\dfrac{\sqrt{3}}{2}e^{-\sqrt{3}\varphi}(E^2+g^{-1}B^2)=0,\nonumber
\eea
where $B^2=g_{xx}(B^x)^2+g_{yy}(B^y)^2+g_{zz}(B^z)^2$ and $E^2=g^{xx}(E_x)^2+g^{yy}(E_y)^2+g^{zz}(E_z)^2$.
\subsection{Wave Solutions}
\label{sec:4}
As previously mentioned, our purpose is to find the effects of higher dimensions on laser interferometers and we should find electromagnetic field in the presence of a scalar field on a GW background. In principle, we should consider the back reaction of all fields on each other and solve equations (\ref{einstein}-\ref{motion,fi}) precisely. This is a hard task in general and so, we will suppose following simplifying assumptions
\begin{enumerate}
\item As before, we will consider $h_{ij}=f_{ij}(z-t), i,j={x,y}$. In other words, we will use (\ref{metric}).
\item We set $f_{\times}=0$ in (\ref{metric}), without loss of generality.
\item Hereafter, we will consider two special relations between electric and magnetic field components namely,
\bea
E_x=B^y,\;\;\;\;\;E_y=-B^x.
\eea
which consistently simplify the computations.
\item We also assume a wavy solution for $\varphi$ in the $z$ direction, i.e. we use the ansatz $\varphi=\varphi(z-t)$ in (\ref{motion,fi}).
\end{enumerate}
Before focusing on explicit solutions of Maxwell equations, let us rewrite the scalar equation (\ref{motion,fi}) and the energy-momentum tensor in (\ref{einstein}) by imposing the above assumptions.

First, it is easy to see that to make $\varphi(z-t)$ be a solution of (\ref{motion,fi}) we should have
\bea
-\frac{1}{4}F_{\mu\nu}F^{\mu\nu}=
%-\frac{1}{2}\left(g^{tt}g^{zz}(E_z)^2+(g^{xx}g^{yy}-(g^{xy})^2)(B^z)^2\right)\nonumber\\
\frac{1}{2}\left((E_z)^2+g^{-1}(B^z)^2\right)=0
\eea
Moreover, independent energy-momentum tensor components would become as
%\begin{widetext}
\bea\label{e-m}
T_{tt}&=&+e^{-\sqrt{3}\varphi}\left( g^{xx}(E_x)^2+g^{yy}(E_y)^2+g^{zz}(E_z)^2+2g^{xy}E_x E_y+\frac{1}{2}g_{tt}\left((E_z)^2+g^{-1}(B^z)^2\right)\right)\nonumber\\
&&-\partial_t\varphi \partial_t\varphi\nonumber\\
T_{tx}&=&-e^{-\sqrt{3}\varphi}\left(g^{yy}E_yB^z +g^{zz}E_xE_z-g^{xy}E_xB^z\right) \nonumber\\
T_{ty}&=&+e^{-\sqrt{3}\varphi}\left(g^{xx}E_xB^z -g^{zz}E_yE_z+g^{xy}E_yB^z\right)\nonumber\\
T_{tz}&=&-e^{-\sqrt{3}\varphi}\left(g^{xx}(E_x)^2+g^{yy}(E_y)^2+2g^{xy}E_x E_y\right)-\partial_t\varphi \partial_z\varphi\nonumber\\
T_{xx}&=&+e^{-\sqrt{3}\varphi}\left(g^{yy}(B^z)^2+\frac{1}{2}g_{xx}\left((E_z)^2+g^{-1}(B^z)^2\right)\right)\nonumber\\
T_{xy}&=&-e^{-\sqrt{3}\varphi}\left(g^{xy}(B^z)^2+\frac{1}{2}g_{xy}\left((E_z)^2+g^{-1}(B^z)^2\right)\right)\nonumber\\
T_{xz}&=&+e^{-\sqrt{3}\varphi}\left(g^{tt}E_xE_z+g^{yy}E_yB^z+g^{xy}E_xB^z\right)\nonumber\\
T_{yy}&=&+e^{-\sqrt{3}\varphi}\left(g^{xx}(B^z)^2+\frac{1}{2}g_{yy}\left((E_z)^2+g^{-1}(B^z)^2\right)\right)\nonumber\\
T_{yz}&=&+e^{-\sqrt{3}\varphi}\left(g^{tt}E_yE_z-g^{xx}E_xB^z-g^{xy}E_yB^z\right)\nonumber\\
T_{zz}&=&+e^{-\sqrt{3}\varphi}\left(g^{xx}(E_x)^2+g^{yy}(E_y)^2+g^{tt}(E_z)^2+2g^{xy}E_x E_y+\frac{1}{2}g_{zz}\left((E_z)^2+g^{-1}(B^z)^2\right)\right)\nonumber\\
&&-\partial_z\varphi \partial_z\varphi\nonumber\\
\eea
%\end{widetext}
%we may also obtain the perturbation metric $h_{\mu\nu}$ up to linear order of other fields $\varphi, F_{\mu\nu}$. With such assumption, one can neglect the energy-momentum tensor of $\varphi$ and electromagnetic fields in (\ref{einstein}) and solve the usual general relativity wave equation. In this case, one have $h_{+}=f_+(z-ct)$ and $h_{\times}=f_{\times}(z-ct)$ where $f_{+,\times}$ are general functions of $z-ct$. We will come back to this case at last section.
%Consequently, using the background metric (\ref{metric}) as a wave propagating in the z direction i.e. $h_{ij}=h_{ij}(z-ct), i,j={x,y}$, we apply it to equations (\ref{gauss}),(\ref{ampere}), (\ref{motion,fi}), in order to find $\varphi, \vec{E}$ and $\vec{B}$ and since we are in linearized regime, we will keep terms up to linear order in $h_{\mu\nu} $.
With a quick look at (\ref{fi,second}) and (\ref{e-m}) it is found that if $f_{\times}=0$ and one turns off the $E_z$ and $B^z$ components of the electromagnetic field then, $\varphi(z-t)$ would be an exact solution of (\ref{fi,second}) and all nonzero components of the energy-momentum tensor would be
\bea\label{Ttt}
T_{tt}&=&T_{zz}=-T_{tz}=\\
&=&e^{-\sqrt{3}\varphi}\left( g^{xx}(E_x)^2+g^{yy}(E_y)^2\right)-\partial_t\varphi \partial_t\varphi\nonumber
\eea
This implies that for making (\ref{metric}) as an exact solution of (\ref{einstein}) , in the absence of $z$ components of electromagnetic fields,  it is enough to set $T_{tt}=0$.

Consequently, presence of $z$ components of electromagnetic field would imply other off-diagonal energy-momentum tensor to be non-zero which means that the plane wave solution (\ref{metric}) is no longer an exact solution of Einstein equation. However, we know that the coefficient of $T_{\mu\nu}$ in (\ref{leinstein}) is of the order of Newton constant which is extremely small. Thus, in the next section, when finding a general solution of Maxwell equation, we will neglect these terms and use the metric (\ref{metric}) as background metric.
\subsubsection{Electric Field in the $x,y$ Directions}
First, we take the components of the electromagnetic field to be in $x,y$ directions and independent of $x,y$ variables i.e. $\vec{E}(z-t)=(E_x, E_y, 0),\;\vec{B}(z-t)=(B_x, B_y, 0)$. Then, it is show that the metric (\ref{metric}) together with
\bea
\varphi(z,t)&=&\varphi(z-t),\nonumber\\
%B_z(z,t)&=&B_0^z,\nonumber\\
%E_z(z,t)&=&\dfrac{E_0^z}{\sqrt{-g}} e^{\sqrt{3}\varphi},\nonumber\\
E_x(z,t)&=&B^y(z,t)=U(z-t),\nonumber\\
E_y(z,t)&=&-B^x(z,t)=V(z-t),
\eea
are the exact solutions of (\ref{einstein}, \ref{maxwell}) and (\ref{motion,fi}) up to the linear order of $h_{\mu\nu}$ provided that (\ref{Ttt}) becomes zero. Here, $U$ and $V$ are the general functions of $z-t$.

Then, the condition $T_{tt}=0$ implies
\bea
e^{\frac{\sqrt{3}}{2}\varphi}=\frac{\sqrt{3}}{2}\int{\sqrt{g^{xx}E_x^2+g^{yy}E_y^2}\;dt}.
\eea
As it is clear, the $x$ and $y$ components of electric and magnetic fields (and the scalar field) propagate wavy.
Notably, this solution  is achieved by considering the independence of the fields on $x$ and $y$ coordinates or equivalently zero-mode $k_x=k_y=0$. Moreover, this solution demonstrates an in-going wave.
However, we will try to find a more general EW solution in which the nonzero modes are taken into account.
\subsubsection{Electric Field in the z-Direction}
Now, let us consider an electric field in the z-direction and take $\Vec{B}=0$. Based on the previous discussion, we know that a non zero $z$ component of electromagnetic field would imply that the plane wave solution of Einstein equation is inexact. But, the perturbation from such $z$ component is very small and we neglect them and we continue the computations using (\ref{metric}).

Then, in accordance with \cite{Cabral:2016klm}, we obtain
\bea\label{em1}
\vec{E}=(0, 0, \frac{E_{0z}}{\sqrt{-g}} e^{\sqrt{3}\varphi}),
\eea
as a simple solution of equations (\ref{amp-gauss}) along with the first and the last equations of (\ref{faraday}). Here, due to the first and the last equation of (\ref{amp-gauss}), $ \vec{E}_0$ can be a general function of $x$ and $y$, but, it is simply found that considering the full Maxwell equations indicates that $\vec{E_0}$ should be a constant.

Thus, it remains to find $\varphi$, i.e. solution of (\ref{fi,second}), by using (\ref{em1}). Considering that we took the metric as $g_{\mu\nu}=g_{\mu\nu}(z-t)$ , as we mentioned earlier, one easily realizes that  equation (\ref{fi,second}) has no analytical solution in the form of $\varphi=\varphi(z-t)$ or $\varphi=\varphi(z+t)$. A possible solution to this problem is to consider the non-zero magnetic field $\vec{B}$.
%Physically, the necessity of the presence of magnetic field comes from the fact that the dynamical scalar field $\varphi$ and non-zero electric field in $z$ direction imply an induced current
%\bea
%J^{\mu}&=&\sqrt{3}(\partial_t\varphi F^{\mu t}+\partial_z\varphi F^{\mu z})\nonumber\\
%&=&\sqrt{3}\partial_t\varphi\left(E_z, -g^{xx}(E_x-B^y), -g^{yy}(E_y+B^x), E_z\right)\nonumber\\
%&=&\sqrt{3}\partial_t\varphi E_z(1,0.0.1).
%\eea
%Therefore, one expect that the nonzero time component of induced current ((t, z)-dependent charge) together with magnetic gauss law implies the existence of non-zero magnetic field in z-direction.
Thus, let us take $E=(0, 0, E_z) $, $ B=(0, 0, B^z)$. Then, (\ref{amp-gauss}) can be solved as
\bea\label{solution}
E_z=\dfrac{E_0^z}{\sqrt{-g}} e^{\sqrt{3}\varphi},\;\;\;\;B^z=B^z(z-t),
\eea
Moreover, from the first and last equations of (\ref{faraday}) we have $\partial_zB^z=\partial_t B^z=0$ and thus
 $B^z$ is a constant $B_0^z$. Finally, equation (\ref{fi,second}) can be solved for $\varphi(z-t)$ as
\bea
e^{\sqrt{3}\varphi}=\frac{B^z_0}{E^z_0}.
\eea
which means that $\varphi$ is also a constant.
%This would result in $E_z(z,t)=\sqrt{-g}B^z_0$.
\subsubsection{Non-zero Mode Wave Solution}
\label{sec:4}
In this section, we would like to find EW solutions which are more general wave solution of previous section i.e. it contains non-zero $k_x$ and $k_y$ modes and non-zero $z$ components of electromagnetic fields. Here, we again consider the conditions $E_x=B^y, E_y=-B^x$ to be held. Due to non-zero $z$ component of electromagnetic field (i.e $E_z$ and $B^z$), the plane wave metric (\ref{metric}) would no longer be an exact solution and that one expects the metric (\ref{metric}) to change. However, hereafter, \textit{we consider the equations of motion of all fields up to the first order of the metric and $E_z$ and $B^z$ components of the electromagnetic field}. In fact, we consider $E_z\ll 1$ and $B^z\ll 1$. Furthermore, such a modification in energy-momentum tensor (\ref{e-m}) is extremely small and so we neglect them and the prodecure is continued using the metric (\ref{metric}). In other words, we consider (\ref{metric}) as background geometry and find the scalar and electromagnetic solutions of the corresponding equation of motion.

Then, the general form of EW equations in the curved space-time is as follows
 \begin{align}
\square F_{ab}=-2R_{acbd}F^{cd}+R_{ae}F^e_ b-R_{be}F^e_ a+J_{a;b}-J_{b;a},
\end{align}
where $J_{a;b}$ is the covariant derivative of current density with respect to $b$. Here, according to new Maxwell equation (\ref{maxwell}), we have
$J^{\mu}=\sqrt{3}\partial_\nu \varphi F^{\mu\nu}$,
then separately writing these equations using GW background metric (\ref{metric}) with $f_{\times}=0$, for electric and magnetic field components, we get the wave equations as
\bea\label{waveeqn2}
&&\partial_t^2 B^z-\partial_z^2 B^z +g^{xx}\partial_x^2 B^z+g^{yy}\partial_y^2 B^z=0,\\ \nonumber \\
&&\partial_t^2 E_z-\partial_z^2 E_z+g^{xx}\partial^2_x E_z+g^{yy}\partial_y^2E_z\nonumber\\
&&\hspace{2.6cm}-\sqrt{3}\partial_t\varphi(\partial_t E_z+\partial_z E_z)=0,\nonumber\\ \nonumber\\
&&\partial_t^2 E_x-\partial_z^2 E_x+g^{xx}\partial^2_x E_x+g^{yy}\partial_y^2 E_x\nonumber\\
&&+\partial_t f_+(\partial_z E_x+\partial_t E_x)
+\partial_t f_{+}(\partial_y B^z -\partial_x E_z)\hspace{2.5cm}\nonumber\\
&&\hspace{4cm}-\sqrt{3}\partial_t\varphi\partial_x E_z=0,\nonumber\\ \nonumber\\
&&\partial_t^2 E_y-\partial_z^2 E_y+g^{xx}\partial^2_x E_y+g^{yy}\partial_y^2 E_y\nonumber\\
&&-\partial_t f_{+}(\partial_z E_y+\partial_t E_y)
+\partial_t f_{+}(\partial_x B^z +\partial_y E_z)\hspace{2.5cm}\nonumber\\
&&\hspace{4cm}-\sqrt{3}\partial_t\varphi\partial_y E_z=0.\nonumber
\eea
It is noteworthy that for obtaining the above equation we used $B^x=-E_y$ and $B^y=E_x$. These assumptions would result that wave equations for $B^{x}, B^{y}, E_{x}$ and $E_{y}$ consistently decouple from each other. Thus, it is enough to find solutions for $B^{z}, E_{x}, E_{y}$ and $E_{z}$.
As it is clear from equations (\ref{waveeqn2}), the equation for $B^z$ is completely decoupled from other components and the equation of $E_z$ is only coupled with the $\varphi$ equation. However, other components have coupled equations with $B^z$ and $E_z$. Consequently, having a solution for $B^z$ and $E_z$ is necessary in order to find solutions for other components.

After all, it is remained to put the solution of $ \varphi $ in the above equations and solve them. 
Due to above assumptions, $E_x=B^y$, $E_y=-B^x$ and $E_z\ll 1, B^z\ll 1$ the last term in (\ref{fi,second}) would be neglected and we can again easily use the solution $\varphi=\varphi(z-t)$ for this equation. We note also that we have $g^{xx}=\frac{1}{g_{xx}}\simeq -(1+f_{+}(z-t))$ and $g^{yy}=\frac{1}{g_{yy}}\simeq -(1-f_{+}(z-t))$  up to the linear order.

Putting these in equations (\ref{waveeqn2}) and using the ansatz
\bea\label{ansatz}
B^i(x,y,z,t)&=&B^i_0e^{i(k.r-\omega t+g_i(u))},\\
E^i(x,y,z,t)&=&E^i_0e^{i(k.r-\omega t+g_i(u))},
\eea
where it was defined $u=z-t$ and $B_0^i, E_0^i, k_x, k_y $ and $ k_z $ are some arbitrary real constants and using the usual dispersion relation
$$\omega^2=k_x^2+k_y^2+k_z^2,$$
then, after a straightforward calculation some details of which are presented in appendix A, a first order differential equation can be found for $g_i(u)$ which can be solved easily.
%the equations (\ref{waveeqn2}) for $\textbf{E}^i(z,t)$ and $\textbf{B}^i(z,t)$ would become
%\bea\label{equations}
%\partial_t^2 \textbf{B}^z &-&\partial_z^2 \textbf{B}^z +\left[b_0+b(z-t)\right] \textbf{B}^z=0,\\
%\nonumber\\
%\partial_t^2 \textbf{E}_z &-&\partial_z^2 \textbf{E}_z +\left[b_0+b(z-t)\right] \textbf{E}_z\nonumber\\
%&-&\sqrt{3}d \sin(z-t)(\partial_t \textbf{E}_z+\partial_z \textbf{E}_z)=0,\nonumber\\
%\nonumber\\
%\partial_t^2 \textbf{E}_x &-&\partial_z^2 \textbf{E}_x +\left[b_0+b(z-t)\right] \textbf{E}_x\nonumber\\
%&+&a \sin(z-t)(\partial_t \textbf{E}_x+\partial_z \textbf{E}_x)\nonumber\\
%&-&i\sin(z-t)(-ak_y\textbf{B}^z+(\sqrt{3}d+a)k_x\textbf{E}_{z})=0,\nonumber\\
%\nonumber\\
%\partial_t^2 \textbf{E}_y&-&\partial_z^2 \textbf{E}_y +\left[b_0+b(z-t)\right] \textbf{E}_y\nonumber\\
%&-&a \sin(z-t)(\partial_t \textbf{E}_y+\partial_z \textbf{E}_y)\nonumber\\
%&-&i\sin(z-t)(-ak_x\textbf{B}^z+(\sqrt{3}d-a)k_y\textbf{E}_{z})=0.\nonumber
%\eea
%where we have defined
%\bea
%b_0&=&k_x^2+k_y^2,\\
%b(z-t)&=&(k_x^2-k_y^2)a \cos(z-t).
%\eea
It is notable that since  $B^z$ and $E_z$ are decoupled from the other field equations, first, we obtain the solution for $B^z$ and $E_z$ straightforwardly and then, insert them into $E_x$ and $E_y$ equations and solve them.

By the above considerations, the final exact solutions of (\ref{waveeqn2}) are as follows
\bea
\label{Bz}
B^z&=&B_0^z e^{-i\left( \omega t-k.r+\theta_0 \right)},\\
\label{Ez}
E_z&=&E_{0z} e^ {-i\left( \omega t-k.r+\theta_0\right)+\frac{\sqrt{3}}{2}\varphi},\\
\label{Ex}
E_x&=&E_{0x}e^{-i\left( \omega t-k.r+\theta_0\right )} \left[e^{-\frac{1}{2}f_+}
+\frac{k_y B_0^z-k_x E_{0z}e^{\frac{\sqrt{3}}{2}\varphi}}{E_{0x}(\omega-k_z)}\right]\nonumber\\
\label{Ey}
E_y&=&E_{0y}e^{ -i\left( \omega t-k.r+\theta_0\right) } \left[e^{+\frac{1}{2}f_+}
-\frac{k_x B_0^z+k_y E_{0z}e^{\frac{\sqrt{3}}{2}\varphi}}{E_{0y}(\omega-k_z)}\right]\nonumber
\eea
where it was defined
\bea
\theta_0=\frac{k_x^2-k_y^2}{2(\omega-k_z)}\int{f_+(u)du}.
\eea
Expectedly, the phase of $B^z$ is changed by the GW but is not affected by the higher dimension (i.e. the field $\varphi$). Instead, $\varphi$ has modified the amplitude of $E_z$ and the Gw has changed the phase again. However, other components are affected by the GW and the,. higher dimension on their amplitude.

For later use, it would be useful to rewrite the above solutions as
\bea
E_i=E_{0i}R_ie^{ -i\left( \omega t-k.r+\theta_0\right)},\;\;\;\;i=x,y,z,
\eea
where
\bea
&&R_z=e^ {+\frac{\sqrt{3}}{2}\varphi},\\
&&R_x=e^{-\frac{1}{2}f_+}
+\frac{k_y B_0^z-k_x E_{0z}e^{\frac{\sqrt{3}}{2}\varphi}}{E_{0x}(\omega-k_z)} \nonumber\\
&&R_{y}=e^{+\frac{1}{2}f_+}
-\frac{k_x B_0^z+k_y E_{0z}e^{\frac{\sqrt{3}}{2}\varphi}}{E_{0y}(\omega-k_z)} .\nonumber
\eea
At the end of this section, it would be interesting to note that the above-mentioned differential equations can have solutions with out-going wave.
\section{Experimental Observation }
\label{sec:5}
\begin{figure}[H]\center
\label{michelson}
\includegraphics[scale=0.7]{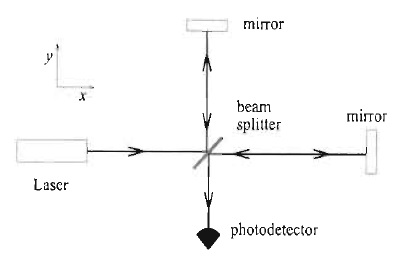}
\caption{A simple Michelson-type interferometer}
\end{figure}

The first direct observation of GWs, GW150914,  was made on 14 September 2015 by LIGO and Virgo collaborations. This observation is based on an interferometer. Here we take a Michelson interferometer for simplicity, figure (\ref{michelson}) in which a monochromatic laser light is sent to a beam-splitter. Due to gravitational waves the travel time of the beams in the two arms will be different, which leads to a different phase and power measured by the photo-detector.
In the previous section, we reached  the electromagnetic fields affected by gravitational waves and one extra dimension. In this section, our aim is to evaluate this extra dimensional effect in a LIGO-like experiment. We expect to see this effect in the photo-detector which shows the total power of light that itself depends squarely on the combined electromagnetic fields, what we will calculate it now.
We will consider expressions (\ref{Ez}) as the electric field of the light, which propagates along the two arms.

First, the travel time in x and y arms is computed by considering the null path, $ ds^2=0 $, for our GW metric. But, before proceeding, note that, in this section, we will return c to the equations and distinguish between GW frequency $ \omega_{gw} $, laser beam frequency $ \omega_{L} $ and the $\varphi$ oscillation frequency $\omega_{\varphi}$ so that at the place $z=0$ we have $f_+(t)=a \cos{(\omega_{gw}t)}$ which is the gravitational wave in the so-called TT gauge and $\varphi(\omega_{\varphi}t)$. Thus, solutions obtained in the previous section would be rewritten using dimensional analysis. For example, one has $\theta_0=\frac{k_x^2-k_y^2}{2(\omega_L-k_z)\omega_{gw}}a \sin(z-t)$.

Moreover, hereafter, we consider $k_x=k_y=k_z=\frac{1}{\sqrt{3}}k_{L}$ and $E_{0i}=\sqrt{c}B_{0}^i$ for $i=x,y,z$ ,without loss of generality.

We will also take $ L_x $ ,$ L_y $, the lengths of x and y arms respectively, as close as possible (which is considered in real interferometers), so that we can replace $ L_x $ and $ L_y $ by $L=(L_x+L_y)/2 $ in the first order terms of the  metric field(parameter $a$) while keeping them in zero order ones by writing $2L_x=2L+\Delta L$ and $ 2L_y=2L-\Delta L$ where $\Delta L=L_x-L_y$.

By Setting the origin of the coordinate system at the beam-splitter (i.e. $r=0$), up to the leading order we get \cite{Husa:2009zz}
\bea\label{t0x}
t^{(x)}_0&=&t-\dfrac{2L_x}{c}+\dfrac{L_x}{c}f(t-\dfrac{L_x}{c}) sinc(\omega_{gw}\dfrac{L_x}{c}),\hspace{1cm}\\
\label{t0y}
t^{(y)}_0&=&t-\dfrac{2L_y}{c}-\dfrac{L_y}{c}f(t-\dfrac{L_y}{c}) sinc(\omega_{gw}\dfrac{L_y}{c}),
\eea
where $t$ is the observation time at which we observe the recombined beam from x and y arms having left the beam-splitter at $ t^{(x)}_0 $, $ t^{(y)}_0 $. We also have $sinc (x)=\frac{\sin x}{x}$ and by $ f(t-\dfrac{L_x}{c})$ we mean the amount of $ f_+ $ at $t= t-\dfrac{L_x}{c}$.

The electric field at the beam-splitter in observation time $t$ is obtained by setting  $t=t_0^{(x)}+\Delta t^{(x)}$ for $x$ arm and $t=t_0^{(y)}+\Delta t^{(y)}$ for $y$ arm in expressions (\ref{Ez}-\ref{Ey}). Here $\Delta t^{(x)}=t-t_0^{(x)}$ and $\Delta t^{(y)}=t-t_0^{y}$ are given by (\ref{t0x}) and (\ref{t0y})  calculated up to the first order of the metric parameter $f_+\varpropto a$. Therefore, in this step,  the solutions are rewritten (\ref{Ez}-\ref{Ey}) up to the first order of $f_+$ and $\varphi$. For this aim, using (\ref{t0x}) and (\ref{t0y}), we have the following expansions
\bea\label{sin}
\sin (t^{(i)})&=&\sin(t_0^{(i)}+\Delta t^{(i)})\\
&\simeq &\sin(t-\frac{2L_{(i)}}{c})+\frac{2L_{(i)}}{c}\cos(t-\frac{2L_{(i)}}{c})+{\cal O}(f_+),\nonumber\\
\cos (t^{(i)})&=&\cos(t_0^{(i)}+\Delta t^{(i)})\nonumber\\
&\simeq &\cos(t-\frac{2L_{(i)}}{c})-\frac{2L_{(i)}}{c}\sin(t-\frac{2L_{(i)}}{c})+{\cal O}(f_+),\nonumber
\eea
where $(i)=x,y$ stand for x and y arms, respectively.

Following the above considerations, one obtains, at the beam-splitter in the observation time $t$, up to the leading order
\bea
\theta_0&=&0,\\
R_z^{(i)}&\simeq & 1+\frac{\sqrt{3}}{2}\varphi(\omega_{\varphi}t^{(i)}),\\
%\theta_z^{(i)}&=&0,\nonumber\\
R_{x}^{(i)}&\simeq &1- \frac{1}{2}f_+(\omega_{gw}t^{(i)}),\\
%\tan(\theta_x^{(i)})&\simeq &\theta_x^{(i)} \simeq -\frac{\sqrt{3}}{2}\frac{\omega_{\varphi}d+\omega_{gw}a}{ck_{L}}\sin(\omega_{\varphi}t^{(i)}),\\
R_{y}^{(i)}&\simeq &1+ \frac{1}{2}f_+(\omega_{gw}t^{(i)}),\\
%\tan(\theta_y^{(i)})&\simeq & \theta_y^{(i)} \simeq  -\frac{\sqrt{3}}{2}\frac{\omega_{\varphi}d-\omega_{gw}a}{ck_{L}}\sin(\omega_{\varphi}t^{(i)}),
\eea
where $\sin(t)$ and $\cos(t)$ are presented by (\ref{sin}).

At the end of the day, we should sum the EWs propagated in x and y arms and recombined in the beam splitter at time $t$. Before that, we should note that reflections and transmission at the mirrors would give overall factors of $-1/2 $ and $+1/2$ for $x$ and $y$ arms, respectively.
The final result would become as
\bea
\vec{E}_{tot}&=&-\frac{1}{2}E_0 e^{-i\omega_L(t-\frac{2L}{c})} \times \hspace{1cm}\nonumber\\
&\times &\LARGE{[}(R_x^{(x)}e^{i(\phi_0+\Delta\phi)}
-R_x^{(y)}e^{i(-\phi_0-\Delta\phi)})\hat{\textbf{i}}\nonumber\\
&+&(R_y^{(x)}e^{i(\phi_0+\Delta\phi)}
-R_y^{(y)}e^{i(-\phi_0-\Delta\phi)})\hat{\textbf{j}}\nonumber\\
&+&R_z^{(x)}e^{i(\phi_0+\Delta\phi)}
-(R_z^{(y)}e^{i(-\phi_0-\Delta\phi)})\hat{\textbf{k}}\Large{]},\nonumber
\eea
where
\bea
\phi_0 &=&\dfrac{\omega_L}{c}(L_x-L_y),\nonumber\\
\Delta\phi(t)&=&-a\frac{\omega_LL}{c}sinc(\frac{\omega_{gw}L}{c})\cos[\omega_{gw}(t-\dfrac{L}{c})],\nonumber
\eea
Furthermore note that similar expression can be written for total magnetic field $\vert B_{tot}\vert$.
%\begin{align}
%\alpha &=\dfrac{E_{0}(d\sqrt{3}+a)-B_0a}{2E_{0}k},\\
%\beta &=\dfrac{E_{0}(d\sqrt{3}-a)-B_0a}{2E_{0}k},
%\end{align}
%Note that we have taken $ k_x=k_y=k $ and $ E_0=E_{0x}=E_{0y}=E_{0z}$, $B_0^x=B_0^y=B_0^z=B_0  $ as these are quantities set by the laboratory.
Consequently, the total observed power at the photo-detector is obtained as
%\begin{widetext}
%\bea\label{newpower}
%&&P \varpropto \left(\vert E_{tot}\vert^2+c^2\vert B_{tot}\vert^2 \right)\\
%&&\simeq E_0^2  [\sin^2(\varphi_0+\Delta\varphi+\Delta\theta_x)+
%\sin^2(\varphi_0++\Delta\varphi+\Delta\theta_y)+
%\sin^2(\varphi_0+\Delta\varphi+\Delta\theta_z)
%+\nonumber\\
%&&\left(\cos(\omega_{\varphi} t^{(x)})+\cos (\omega_{\varphi} t^{(y)})\right)
%\left(\frac{1}{4}a\sin^2(\varphi_0+\Delta\varphi+\Delta\theta_x)-
%\frac{1}{4}a\sin^2(\varphi_0++\Delta\varphi+\Delta\theta_y)+
%\frac{\sqrt{3}}{2}d\sin^2(\varphi_0+\Delta\varphi+\Delta\theta_z) \right)] ,\nonumber
%\eea
%\end{widetext}
%\begin{widetext}
\bea\label{newpower}
&&P\varpropto \left(\vert E_{tot}\vert^2+c^2\vert B_{tot}\vert^2 \right)\\
&&\simeq 2E_0^2\left(3+\frac{\sqrt{3}}{2}\left(\varphi(\omega_{\varphi}t^{(x)})
+\varphi(\omega_{\varphi}t^{(y)})\right)\right)\sin^2{\left(\phi_0+\Delta\phi\right)}\nonumber
\eea
%\end{widetext}
where
%\bea
%\Delta\theta_x&=&\frac{1}{2}(\theta_x^{(x)}-\theta_x^{(y)})\\
%&=&\frac{\sqrt{3}d}{4}\frac{\omega_{\varphi}}{ck_{L}} \left(\sin{(\omega_{\varphi}t^{(x)})}-\sin{ (\omega_{\varphi}t^{(y)})}\right)\nonumber\\
%&\simeq &\sqrt{3}d\;\frac{\omega_{\varphi}^3}{\omega_{L}}\frac{L\Delta L}{ c^2} sinc\left(\frac{\omega_{\varphi}\Delta L}{c}\right)\sin{\left[\omega_{\varphi}(t-\frac{2L}{c})\right]}\nonumber\\
%&+&{\cal O}(d^2),\nonumber\\
%\Delta\theta_y&=&\frac{1}{2}(\theta_y^{(x)}-\theta_y^{(y)})\\
%&=&\frac{\sqrt{3}d+a}{4}\frac{\omega_{\varphi}}{ck_{L}} \left(\sin{(\omega_{\varphi}t^{(x)})}-\sin{ (\omega_{\varphi}t^{(y)})}\right)\nonumber\\
%&\simeq &(\sqrt{3}d+a)\;\frac{\omega_{\varphi}^3}{\omega_{L}}\frac{L\Delta L}{ c^2} sinc\left(\frac{\omega_{\varphi}\Delta L}{c}\right)\sin{\left[\omega_{\varphi}(t-\frac{2L}{c})\right]}\nonumber\\
%&+&{\cal O}(d^2),\nonumber\\
%\Delta\theta_z&=&0,
%\eea
%and
%\bea
%&&\cos(\omega_{\varphi} t^{(x)})+\cos (\omega_{\varphi} t^{(y)})\simeq \\
%&&\simeq  2\cos{(\omega_{\varphi}(t-\frac{2L}{c}))}-
%\frac{4\omega_{\varphi}L}{c}\sin{(\omega_{\varphi}(t-\frac{2L}{c}))}.\nonumber
%\eea
\bea
&&\varphi(\omega_{\varphi} t^{(x)})+\varphi (\omega_{\varphi} t^{(y)})\simeq \\
&&\simeq  2\varphi{(\omega_{\varphi}(t-\frac{2L}{c}))}+
\frac{4L}{c}\frac{d}{dt}\varphi{(\omega_{\varphi}(t-\frac{2L}{c}))}.\nonumber
\eea
%In formula (\ref{newpower}), in calculation of coefficients of  $\sin^2(\phi_0+\Delta\phi$\\
%$\Delta\theta)$ and $\sin^2\left(\phi_0+\Delta\phi\right)$, we neglected terms of order of $a^2, d^2$ and $ad$.
To discuss our result (\ref{newpower}) some comments should be noted as follows:
\begin{itemize}
\item As shown in (\ref{newpower}), the GWs add a term to the phase due to $\Delta\phi$ while the extra dimension shows itself in changing the amplitude of $P$ by term $\frac{\sqrt{3}}{2}(\varphi(\omega_{\varphi} t^{(x)})+\varphi (\omega_{\varphi} t^{(y)}))$.
\item In formula (\ref{newpower}), we have three \textbf{\textsl{known-controllable}} parameters $\omega_L, L$ and $\Delta L$. The parameter $\omega_L$ is adjustable from the experimenter. For parameter $L$,
in a typical Michelson interferometer, one maximizes the phase $\Delta\phi_{Mich}=2\Delta \phi$ and the optimal length of the arms is obtained by imposing $\frac{\omega_{gw}L}{c}=\frac{\pi}{2}$ implying that $L\simeq750km\frac{100Hz}{f_{gw}}$ where $f_{gw}=\frac{\omega_{gw}}{2\pi}$. Such hundreds of kms arms are reduced to a few kms using the Fabry-Perot interferometer \cite{Husa:2009zz}.
%But, for $\Delta L=L_x-L_y$, one would like to take $L_x$ and $L_y$ as close as possible but we note that the phase shift $\Delta \theta$ is proportional to $\Delta L$. Therefore, a difference between arms may show a signal of extra dimension!
\item In formula (\ref{newpower}), we have two \textbf{\textsl{known-uncontrollable}} parameters of $f_+\propto a$ and $\omega_{gw}$. Due to the direct detection of GWs by LIGO experiment the source of which was the merger of  black holes \cite{Abbott:2016blz}, we have $\omega_{gw}=35Hz$ to $250Hz$ and $a=10^{-21}$ (strain peak). Nevertheless, these parameters clearly vary for different astrophysical phenomenons.
\item In formula (\ref{newpower}), we have two \textbf{\textsl{unknown-uncontrollable}} parameters $\varphi$ and $\omega_{\varphi}$. The oscillation frequency of the fifth dimension, if it oscillates anyway, is quite unknown. However for parameter $\varphi$, recall that it is related to the radius of the fifth dimension by (\ref{R}). Up to now, we have no higher dimensional fundamental theory confirmed by any experiment. However, up to the theoretical point of view, different proposed higher dimensional theories predict different sizes for extra dimensions. It starts from the extra dimension at millimeter \cite{ArkaniHamed:1998rs} up to the higher dimension of order of string scale.
\end{itemize}
At the end of this section, it is notable that in our toy model the radius of the extra dimension is related to $\varphi$ as $\varphi\propto R_0^2$ and that for theories with extra dimension at Millimetre or smaller than it we have $\varphi< 10^{-6}$. So that, due to (\ref{newpower}), its effect on power signal in an interferometer is very weak and that its detection in prsent gravitational based experiments is very hard.
%Remembering all the above comments in mind, one realizes that the power $P$ would be affected by higher dimension due to the parameters $d, \frac{\omega_{\varphi}L}{c}, \frac{\omega_{\varphi}}{\omega_{gw}}$ and the ratio of coefficients of $\Delta{\theta}$ and $\Delta{\phi}$ which up to some constant is equal to $\Delta=\frac{d}{a}\frac{\omega_{\varphi}^3}{\omega_{L}^2}\frac{\Delta L}{c}$.
\section{ Summary and conclusion }
\label{sec:6}
Due to the detection of GWs, a highly important question is whether higher dimensional effects would appear in gravitational wave physics. In this paper, we considered a five dimensional pure gravity theory compactified on a circle. We studied this four dimensional minimally interacting scalar-Maxwell theory on the background of a GW.

First, by varying the action, the equations of motion for scalar, electric and magnetic fields were achieved on a GW background and then, we obtained the explicit wave solutions for these fields.

Next, we considered the recombination of EWs in a typical Michelson interferometer. In particular, we supposed a general fifth dimensional field and obtained, up to the first order, the change of the amplitude of EWs due to the presence of a scalar field in terms of the scalar, electromagnetic and GW parameters. In particular, we obtained  the power $P$ affected by a higher dimension
%due to the parameters $d, \frac{\omega_{\varphi}L}{c}, \frac{\omega_{\varphi}}{\omega_{gw}}$ and the ratio of coefficients of $\Delta{\theta}$ and $\Delta{\phi}$ which up to some constant is equal to $\Delta=\frac{d}{a}\frac{\omega_{\varphi}^3}{\omega_{L}^2}\frac{\Delta L}{c}$.
This change in the amplitude of power may be an observable quantity and its detectability would depend on the accuracy of future GW interferometers.

Eventually, we would like to point out that, in this paper, we have studied a toy model  to demonstrate the effect of one extra dimension on a LIGO-like experiment. However, it would be highly important to consider theories with more than one extra dimension such as theories  from string compactification or higher dimensional theories with higher curvature terms and the like.
%%%%%%%%%%%%%%%%%%%%%%%%%%%%%%%%%%%%%%%%%%%%%%%%%%%%%%%%%%%%%%%%%%%%%%%%%%%%%
\section{Acknowledgment}
The authors would like to thank Dr. Ali Hasanbeigi and Dr. Nahid Ahmadi for useful comments.

Mohammad A. Ganjali would like to thank the Kharazmi university
for supporting the paper with grant.
%%%%%%%%%%%%%%%%%%%%%%%%%%%%%%%%%%%%%%%%%%%%%%%%%%%%%%%%%%%%%%%%%%%%%%%%%%%%%%%%
\section{Appendix}
In this appendix, we present some detail calculations of solution of equations (\ref{waveeqn2}). Let us consider the $E_x$ equation. Procedure for other fields is quite similar. Using the ansatz (\ref{ansatz}) and dispersion relation $\omega^2=|k|^2$, one find the following equation for $E_x$
\bea
&&-2(\omega-k_z)\frac{dg}{du}E_x+\left(-i(\omega-k_z)\partial_t f_++(k_x^2-k_y^2)f_+\right)E_x+\nonumber\\
&&+ik_y\partial_t f_+B^z-ik_x(\partial_t f_++\sqrt{3}\partial_t \varphi)E_z=0.
\eea
Then, using the solutions of $E_z$ and $B_z$ and defining new function
$$G(u)=e^{\tilde{g}(u)}=e^{g(u)-\theta_0}$$ one obtains
\bea
\frac{dG}{du}&+&\frac{1}{2}\partial_uf_+ G=\\
&&\frac{k_yB_0^z\partial_u f_+-k_xE_0^z(\partial_u f_++\sqrt{3}\partial_u \varphi)e^{\frac{\sqrt{3}}{2}\varphi}}{2(\omega-k_z)},\nonumber
\eea
which is a first order differential equation in standard form $\frac{dG}{du}+P(u)G=Q(u)$.

\bibliographystyle{}
 \bibliography{}

\end{document}